# Isotropic conduction and negative photoconduction in ultrathin PtSe$_2$ films

**Francesca Urban**[1,2,3], **Farzan Gity**[4], **Paul K. Hurley**[4], **Niall McEvoy**[5], **and Antonio Di Bartolomeo**[1,2,*]

[1] Physics Department, University of Salerno, via Giovanni Paolo II, 132, 84084 Fisciano, Salerno, Italy

[2] CNR-Spin, via Giovanni Paolo II, 132, 84084 Fisciano, Salerno, Italy

[3] INFN-Gruppo collegato di Salerno, via Giovanni Paolo II, 132, 84084 Fisciano, Salerno, Italy

[4] Tyndall National Institute, University College Cork, Cork, Ireland

[5] AMBER & School of Chemistry, Trinity College Dublin, Dublin 2, Ireland

* Correspondence: adibartolomeo@unisa.it

**Abstract:** PtSe$_2$ ultrathin films are used as the channel of back-gated field-effect transistors (FETs) that are investigated at different temperatures and under super-continuous white laser irradiation. The temperature-dependent behavior confirms the semiconducting nature of multilayer PtSe$_2$, with p-type conduction, a hole field-effect mobility up to 40 $cm^2V^{-1}s^{-1}$ and significant gate modulation. Electrical conduction measured along different directions shows isotropic transport. A reduction of PtSe$_2$ channel conductance is observed under exposure to light. Such a negative photoconductivity is explained by a photogating effect caused by photo-charge accumulation in SiO$_2$ and at the Si/SiO$_2$ interface.

**Keywords:** PtSe$_2$; field effect transistor; laser irradiation; electrical conduction; temperature; negative photoconductivity;

The family of transition-metal dichalcogenides (TMDs), such as the most known MoS$_2$, WSe$_2$, WS$_2$, has been heavily investigated over the last decade due to their intriguing and layer-tunable properties combined with their ease of fabrication [1–7]. The bulk materials, made of atomic layers held together by van der Waals forces, can be easily exfoliated to obtain single- or few-layer nanosheets [8,9] and the electrical and optical properties of these materials are strongly dependent on their thickness [10,11]. Indeed, the modulation of bandgap via changing the number of layers enables the use of TMDs as field-effect transistor (FET) channels in optical sensors with high photo-response [12,13]. Moreover, the intrinsic n- or p-type doping is useful for the construction of p-n heterojunctions [14–16].

To date, in addition to the exfoliation methods, various scalable and controllable growth techniques, mostly based on chemical vapor deposition (CVD) and laser ablation, have been developed, enabling the synthesis of large-area flakes with fine thickness control and consequent tailoring of the chemical, optical and electrical properties [17–19].

Different from MoS$_2$, WSe$_2$ or other group-6 TMDs, dichalchogenides based on group-10 transition metals have just recently gained popularity and not been fully explored yet.

Platinum diselenide (PtSe$_2$) and ditelluride (PtTe$_2$), along with their palladium analogues, were theoretically predicted in the sixties of the last century but have only recently been isolated and investigated as 2D materials [20–26]. Numerical calculations of their electronic structure and properties encouraged their use in electronic applications. These materials crystallize in an octahedral lattice structure where the transition metal atoms are coordinated with six chalcogens. Each layer is a two-dimensional packed array of metal atoms sandwiched between two similar arrays of chalcogens, and bonded by van der Waals forces to form the multilayer structure. The main difference between TMDs of group 6 and the group 10 is that the presence of d-electrons in the group-10 transition metals gives rise to additional semiconductor bands [20].

PtSe$_2$, the material under study in this paper, is a semimetal in bulk form, with slightly indirect overlap of the conduction and valence bands, that undergoes a semimetal-to-semiconductor transition when it is thinned to a few atomic layers [27–29]. Monolayer PtSe$_2$ has an indirect bandgap of ~1.2 eV, which is expected to reduce to 0.3 eV for the bilayer [22,30]. The variability of electrical properties, combined with environmental stability, is attracting growing attention from both fundamental and application standpoints. For instance, the bandgap covers the spectral range that is important for telecommunications and solar energy harvesting [31], and the carrier mobility (theoretically predicted up to 4000 $cm^2V^{-1}s^{-1}$ [32] and experimentally found to be around 200 $cm^2V^{-1}s^{-1}$ [33,34]), competitive with black phosphorus, can enable fast electronic devices [33,35–37]. Other interesting properties include the catalytic activity [38,39] and the sensitivity to analytes such as NO$_2$, NO, NH$_3$ or ethanol [40,41]. Furthermore, the direct selenization of platinum films on a chosen substrate, and the rather low temperatures (400 °C) required for the thermally assisted synthesis of PtSe$_2$, might give this material a boost for integration in semiconductor technologies for mass production [21,22,28].

In this work, we study the electrical conduction in multilayer PtSe$_2$ sheets along orthogonal directions using back-gated field-effect transistors. The electrical behavior is analyzed over a wide temperature range demonstrating semiconducting p-type conduction, with reasonable gate modulation and relatively high hole mobility. The electrical conduction along orthogonal directions does not show any significant difference and the intrinsic p-type behavior of the transistors is stable under air exposure. Laser irradiation reveals more complex phenomena than the usual current increase by photo-generation. Indeed, excitation by laser pulses unexpectedly reduces the channel conductance. We explain such a negative photoconductivity (NPC) as occurring due to a photogating effect, which arises from photocharge accumulation in the SiO$_2$ dielectric and at the Si/SiO$_2$ interface.

The devices were fabricated over Si/SiO$_2$ substrate (85 nm thermally grown oxide on p-type silicon, $\rho \sim 0.001 - 0.005\ \Omega cm$). The PtSe$_2$ film on SiO$_2$ was obtained by direct selenization of a previously sputtered Pt film (nominal thickness 0.7 nm) following procedures described elsewhere [40]. Briefly, Pt films were placed in the center of the downstream zone of a two-zone furnace where they were heated to 400 °C. Se pellets were independently heated to 220 °C in the upstream zone and 150 sccm of Ar:H$_2$ (90%:10%) carried Se vapor from the upstream to the downstream zone, where it reacted with the Pt films. A growth time of 2 hours ensured that Pt was completely converted to PtSe$_2$. During this process the film thickness increases by a factor of ~3.5 – 4 [21].

The PtSe$_2$ film was transferred using a polymer-based process from the growth substrate to a fresh Si/SiO$_2$ substrate [29], and it was then patterned using photoresist masking and a SF$_6$-based inductively coupled plasma (ICP) etching process. This was followed by patterning Ni:Au (20 nm : 150 nm) metal contacts using a standard lift-off process.

The as-synthesized PtSe$_2$ film has a 1T crystal structure, where six selenium atoms are bonded to a platinum atom located at the center of an octahedral geometry, as shown in Figure 1(a)[42]. The lattice in a top view (see Figure 1(b)) appears as two hexagonal arrays, shifted with respect to each other, with Pt (yellow spheres) and Se (blue spheres) atoms centering the basis [40,43].

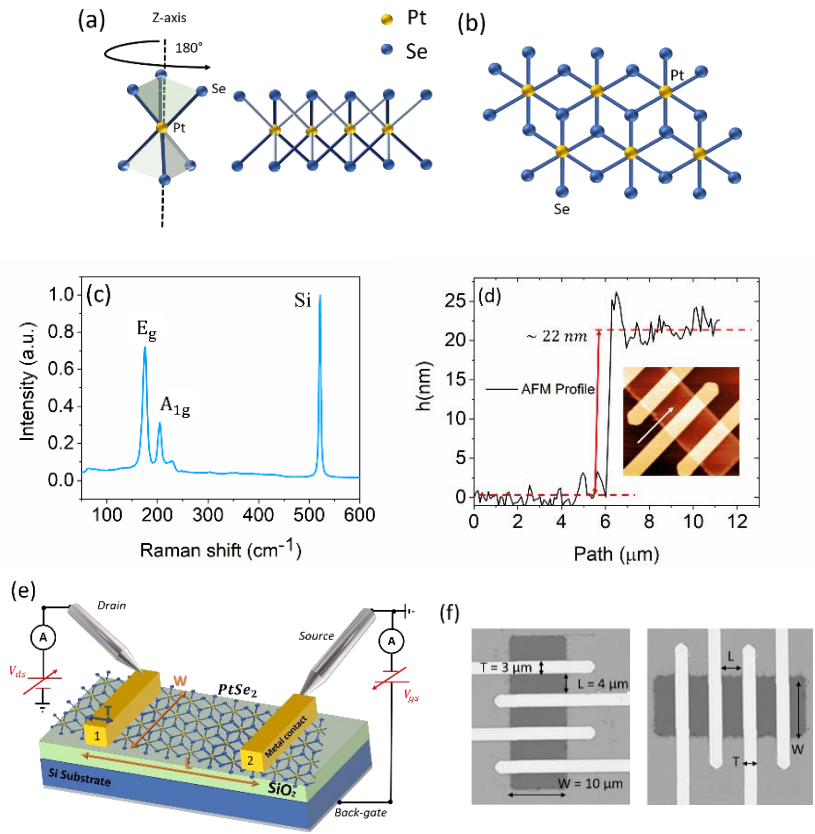

**Figure 1.** (**a**) Scheme of the octahedral basis and layered structure of PtSe$_2$ layer and (**b**) its hexagonal top viewed array. (**c**) Raman characterization of the prepared PtSe$_2$ sheet, showing the two PtSe$_2$ peaks at 176 cm$^{-1}$ and 205 cm$^{-1}$, and the Si peak. (**d**) AFM image and step height profile. (**e**) Schematic of the measurement setup. (**f**) Optical images of two selected devices in the horizontal and vertical configuration.

Raman spectroscopy was used to characterize the as-synthesized films. The characteristic $E_g$ (~$176\ cm^{-1}$) and $A_{1g}$ (~$205\ cm^{-1}$) modes (see Figure 1(c)) were observed confirming the successful synthesis of PtSe$_2$. The position and relative intensity of these modes is consistent with the synthesis of multilayer PtSe$_2$ [43]. The step height measured by AFM (Figure 1(d)) is ~ $22\ nm$ and includes the $19\ nm$ SiO$_2$ over-etch applied during the PtSe$_2$ film patterning. Therefore, the thickness of PtSe$_2$ is ~3 $nm$, corresponding to about 6 layers (PtSe$_2$ monolayer is 0.5 nm thick [44]).

The devices were measured in two- and four-probe configurations in a Janis Probe Station (Janis ST-500 probe station) equipped with four nanoprobes connected to a Keithley 4200 SCS (semiconductor characterization system), under different conditions. A scheme of the contacted back-gated device is shown in Figure 1(e): The metal contacts were used as the drain and source electrodes while the probe station chuck, connected to the silicon substrate, provided the gate voltage.

The following analysis was conducted on two main types of devices, having the same channel length (L) and width (W) and contact leads (T) in the direction of the channel. The devices were fabricated from the same PtSe$_2$ sheet, simultaneously patterned to form a horizontal and a vertical channel, an example is displayed by the optical microscope images of Figure 1(f). Identical contacts, differing only in orientation, allow the measurement of the channel conductance in the two perpendicular directions.

The electrical measurements were performed at a constant air pressure of $10^{-3}$ mbar and at different temperatures, as well as in dark conditions and under irradiation from a super-continuous laser source (NKT Photonics, Super Compact, wavelength ranging from 450 nm to 2400 nm, at 30 mW/cm$^2$).

Using the two-probe configuration, we studied the variation of the PtSe$_2$ conductance G as a function of temperature T from 400 K to near liquid-nitrogen temperature. The G-T curves for the horizontal and vertical samples, reported in Figure 2(a), show similar trends with the channel conductance decreasing when T is slowly cooled down from 400 K to 90 K. The G-T behavior reveals the semiconducting nature of PtSe$_2$ ultrathin films.

The I-V current-voltage characteristics, at zero gate voltage, in two- and four-probe configurations, are reported in the inset of Figure 2(a), for both the two chosen horizontal- and the vertical-channel devices. The comparison of two- and four-probe measurements, with no appreciable difference, confirms the negligible effect of the metal contacts on the channel conductance as well as the ohmic behavior of the semiconductor/metal junctions, enabling the use of the two-probe configuration for further analysis.

To better understand the semiconducting nature of the material, we performed transfer characteristic measurements at two extreme temperatures, 290 K and 90 K. The gate voltage was intentionally limited between -10 and 10 V to avoid the breakdown of the 85 nm gate oxide. Figure 2(b) shows decreasing conductance for positive gate voltage, indicative of p-type intrinsic channel doping, and a modulation of a factor 2 in the narrow explored gate voltage range, consistent with the narrow bandgap of multilayer PtSe$_2$.

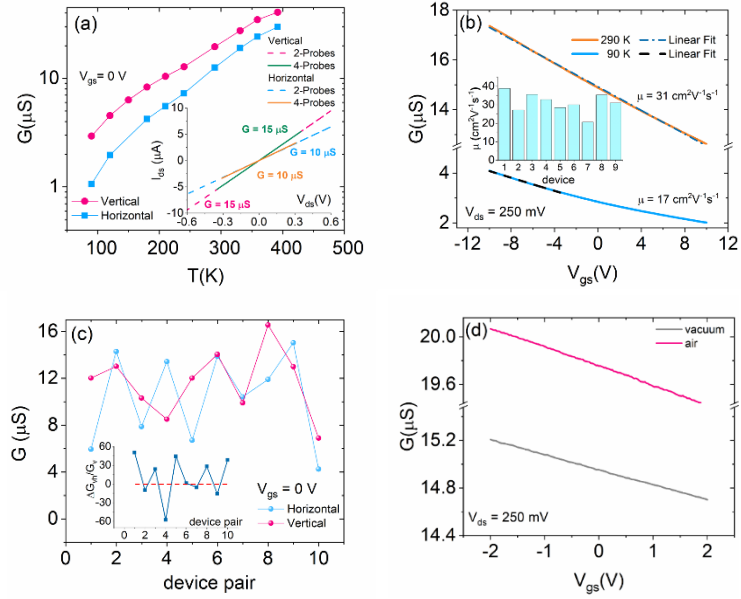

**Figure 2.** (**a**) G vs T curves for horizontal (cyan) and vertical (magenta) devices on semilogarithmic scale, at $V_{gs} = 0\,V$; the inset shows the I-V characteristics for horizontal and vertical device in 2- (4-) probe configuration plotted in cyan dashed (orange straight) line and in magenta dashed (green straight) line, respectively. (**b**) Transfer characteristics at 90 K (cyan curve) and 290 K (orange) for the vertical device with the linear fits (black and blue dashed curves respectively) used to estimate the maximum hole mobility. The inset reports the hole mobility for different devices in an analog configuration. (**c**) Room temperature conductance for horizontal (cyan) and vertical (magenta) device pairs. Each pair is formed by a horizontal and a vertical device fabricated next to one another to avoid potential impact of thickness. The inset shows the percentile variation $\frac{\Delta G_{vh}}{G_v} = \frac{G_v - G_h}{G_v}$ between the two conductances for each device pair. (**d**) Transfer characteristics in vacuum $\sim 10^{-3}$ mbar (grey curve) and air (magenta curve), for the vertical device.

From the transfer characteristics, we evaluated the field effect mobility as $\mu = \frac{L}{W C_{ox} V_{ds}} \frac{dI_{ds}}{dV_{gs}}$ ($V_{ds}$ is the source-drain bias and $C_{ox} = 3.11\,nFcm^{-2}$ is the SiO$_2$ capacitance per area), finding values in the range $15 - 40\,cm^2V^{-1}s^{-1}$ at room temperature (as reported in the inset of Figure 2(b) for for the vertical devices) and $10 - 20\,cm^2V^{-1}s^{-1}$ at 90 K. The temperature dependence of the mobility is ascribed to Coulomb scattering due to fixed charges [6,45].

These values can be considered rather high if compared to the mobilities commonly reported for similar PtSe$_2$ or other TMD-based devices [5,24,34,46–48].

The electrical conductance of several pairs of horizontal and vertical devices, at room temperature and grounded gate, has been investigated and it is reported in Figure 2(c). The device pairs were chosen as close as possible to each other (few $\mu m$) to avoid potential impact of thickness between distant zones of the wafer. The plot shows that the conductance of horizontal and vertical devices have fluctuations up to 60% but originate distributions that are consistent with each other. Otherwise stated, Figure 2(c) demonstrates that conduction of PtSe$_2$ films is isotropic. The isotropic conduction is consistent with the observation of randomly-oriented grains in ultrathin PtSe$_2$ films reported elsewhere [40,43,49].

Finally, Figure 2(d) shows that passing from vacuum (∼ 1 mbar) to ambient pressure does not change the channel doping from p-type to n-type as observed with other TMD materials, strongly influenced by the environmental atmosphere [6,24,50,51]. The p-type behavior of $PtSe_2$ is preserved when changing either the chamber temperature or pressure, and the conduction enhancement in air is attributed to the p-doping of oxygen molecules, which are electron acceptors.

To further investigate the behavior of $PtSe_2$, we measured the electrical conduction of the device under pulsed laser irradiation, both in vacuum ($10^{-3}$ mbar) and at room pressure.

Figure 3(a-b) show the device channel current under super-continuous laser irradiation, with the light switched on and off every two minutes for 12 cycles. The experiment was conducted with the laser source at 30 mW/cm². After the on/off cycles, the laser is turned off and the device slowly returns to the initial conduction state. Surprisingly, each laser irradiation (on-pulse) induces a reduction of the current. This behavior, that is referred to negative photoconductivity, is opposite to the current increase normally observed under light as an effect of photo-generation [52–55].

It is important to note that the laser beam induces reversible changes in the irradiated device. The current lowering lasts for different periods of time, ranging from minutes to several hours. The comparison of Figure 3(a) and 3(b) shows that the restoration of the initial state is significantly faster in air at room pressure.

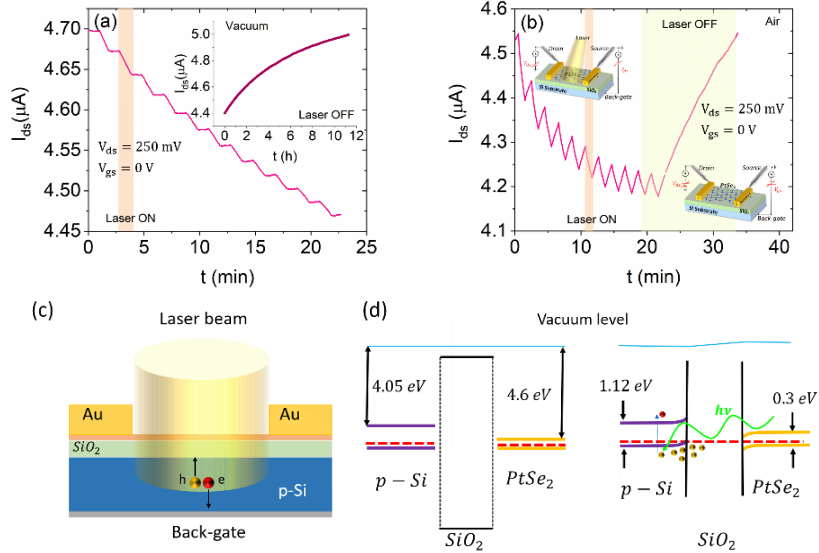

**Figure 3.** $I_{ds}$ vs time characteristics under super-continuous laser beam irradiation, wavelength ranging from 450 nm to 2400 nm, at $10^{-3}$ mbar (**a**) and room pressure (**b**). The inset of panel (**a**) shows the recovery of the device current after several hours from the laser irradiation. (**c**) Schematic of the device under laser irradiation and (**d**) band diagram of the p-Si/$SiO_2$/$PtSe_2$ structure and charge accumulation under laser pulse.

The NPC mechanism can be explained by considering the dominance of the photogating effect caused by the electron-hole pair photogeneration in Si or $PtSe_2$ close to the $SiO_2$ interface [56]. When the device is irradiated, the incident laser is mostly absorbed by the Si substrate [57] leading to the creation of electron-hole pairs in the substrate. While the generated electrons are swept to ground, the photo-generated holes accumulate at the Si/$SiO_2$ interface or in $SiO_2$ trap states (see Figure 3(c)), favored by the vertical up-bending of the Si bands at the Si/SiO$_2$ interface due to the different electron affinity of Si ($\sim 4.05\ eV$) and PtSe$_2$ ($\sim 4.6\ eV$), Figure 3(d).

The so-accumulated positive charge below the PtSe$_2$ channel plays the role of a positive gate voltage (photogating effect) which biases the device and decreases the channel current, according to the p-type behavior of the transistor.

The restoration of the initial state, in vacuum, is slower due to the limited diffusion of the positive photo-charge accumulated at the at the Si/SiO$_2$ interface or in the SiO$_2$ layer. It becomes faster in air because of the counterdoping effect of adsorbates, such as oxygen or water, which increase the p-doping of the PtSe$_2$ channel.

In conclusion, we investigated the electrical transport in PtSe$_2$ films along perpendicular directions. We used PtSe$_2$ as the channel material of field effect transistors in back-gated configuration. The conductance vs temperature behavior confirmed the semiconducting nature of ultrathin PtSe$_2$ films and the transistor characterization indicated intrinsic p-type conduction. The silicon back-gate enabled channel current modulation with field effect mobility up to $\sim 40\ cm^2 V^{-1} s^{-1}$ at room temperature. The comparison of the channel conductance along perpendicular directions for a large number of devices led to the conclusion that the conduction is isotropic being mainly dominated by the polycrystalline structure of the PtSe$_2$ film. Finally, we found a negative photoconductivity under laser irradiation, indicating a dominant photogating effect.

**Acknowledgments:** This work is partially supported by Science Foundation Ireland (SFI) through grants 15/IA/3131 and 12/RC/2278 and 15/SIRG/3329, 12/RC/2278_P2 and by Italian Ministry of University and Research (MIUR), projects Pico & Pro ARS01_01061 and RINASCIMENTO ARS01_01088.

**Data availability statement:** The data that support the findings of this study are available from the corresponding author upon reasonable request.